\documentclass[10pt]{elsart}
\usepackage{amssymb,amsmath}
\usepackage{graphics,graphicx}
\usepackage{epsfig,epic,eepic,psfig,rotate,float}
\usepackage{subfigure}

\begin{document}

\begin{frontmatter}



\title{Cluster size distribution of infection in a
system of mobile agents}


\author[a1]{M.C. Gonz\'alez\corauthref{cor1}}
\corauth[cor1]{Tel.: +49-711-685-3594; fax: +49-711-6853658.}\ead{marta@ica1.uni-stuttgart.de}
\author[a1,a2]{\underline{H. J. Herrmann}}
\author[a1,a2]{A. D. Ara\'ujo}
\address[a1]{Institute for Computational Physics, University of 
Stuttgart, Pfaffenwaldring 27, 70569 Stuttgart, Germany}
\address[a2]{Departamento de F\'{\i}sica, Universidade Federal do
             Cear\'a, 60451-970 Fortaleza, Brazil}
\begin{abstract}
Clusters of infected individuals are defined on data from health 
laboratories, but this quantity has not been defined and characterized
by epidemy models on statistical physics. For a system of mobile
agents we simulate a model of infection without immunization and 
show that all the moments of the cluster size distribution at the critical 
rate of infection are characterized by only one exponent, 
which is the same exponent that determines the behavior of the total number 
of infected agents. No giant cluster survives independent on 
the magnitude of the rate of infection.
\end{abstract}
\end{frontmatter}

\section{Introduction}
In a small and highly urbanized nation like Singapore
dengue outbreaks or epidemics are identified as 
``clusters''. A dengue {\bf cluster} or focus of transmission is defined
as at least two confirmed cases, with no recent travel
history, that are located within 200 m  of each other
(taken as the flight range of the {\em Aedes aegypti})
and whose dates of the onset of symptoms are within three
weeks of each other~\cite{dengue}. 
Some efforts have been directed 
towards the characterization
of 'SIS' models of infections, or epidemics without 
immunization~\cite{Mollison,Grassberger,satorras}, that is the state
of the particles are healthy or infected, and are susceptible
to re-infection after healing, thus the name of the model
(SIS: susceptible-infected-susceptible). 
Analytical and numerical expressions describe the dynamics of 
the $SIS $ model in terms of the rate of spreading $\lambda$,
the evolution of the survival probability of infection $P(t)$, the 
mean number of infected agents $n(t)$ and the mean square distance of 
spreading $R^{2}(t)$ in time, which are quantities difficult to
compare with real data of epidemics. This work suggests an application 
of potential comparison with public health data, analyzing a scaling 
function for {\bf clusters numbers} on  a $SIS$ model of infection.\\
The second important ingredient of this work is the
mobility of agents, contrasted to most of the models of 
epidemy where the population is modeled 
by static networks~\cite{Mollison,Grassberger,satorras}.
In a previous work~\cite{us} we showed that the $SIS$ model of infection
on a system of mobile agents has critical exponents which
depends on the density of the system, i.e spatial correlations
and mobility of the agents play an important role. We obtained a crossover
from mean field behavior for low densities to static $2D$-lattices
for higher densities. Here we use our model of mobile agents to define
clusters of infections and analyze its dependency on the rate 
of infection $\lambda$ (defined in detail bellow) and on the mobility
of the agents.\\ 
We propose a time-evolving network model: A link between 
two moving  agents is created when they collide with 
each other and there is transmission of the infection among them 
(i.e through infected-susceptible interactions), the
link lasts a characteristic time of infection.\\ 
We find that the network of clusters of infections
remains disconnected and no matter how large the rate of infection,
no giant cluster is formed. We show that in the transition
to spreading,  the moments of the cluster size distribution are described by 
an exponent $\beta$, which is the exponent that characterizes the fraction 
of infected mass $F_{IM}=N_{Inf}/N$, defined as 
the ratio of the number of infected agents ($N_{Inf}$) and
the total amount of population ($N$). Thus the number of clusters 
depends on $\lambda$, and mobility and spatial correlation of 
the agents influence its dependency.
\section{Model}
  $N$ soft disks, with radius $r_{0}=1$, represent
agents which move in a two dimensional cell of linear size
$L$, with density $\rho=N/L^{2}$. The system has periodic 
boundary conditions and is initialized as follows: the agents are 
placed in the cell with the same velocity modulus
$v$ and randomly distributed directions, positions and 
states: 'infected' or 'susceptible'. If a susceptible agent $i$ collides
with an infected agent $j$ (i.e $|\mbox{\boldmath$r$}_{i}-\mbox{\boldmath$r$}{j}|<=2r_{0}$), 
then $i$ becomes infected. Each infected agent heals and becomes
susceptible again after a fixed number of time steps, called
the 'time of infection' ($\Delta t_{inf}$), which is a free parameter
of the model.\\
The physical interaction of the agents is modeled by molecular dynamic
with a leap-frog integration method \cite{Rapaport}, the interaction potential
is a $12-6$ Lennard-Jones truncated potential (see more details in \cite{us}).\\
The resulting model is a contact process~\cite{Dickman}, where 
the infected species become extinct unless the infection spreads rapidly
enough. The transition between survival and extinction depends on a critical
rate of spreading $\lambda_{c}$ that marks the transition into an absorbing
state. The infection rate $\lambda$ is defined as the number of agents 
one agent infects before healing. For this system,
\begin{equation}
\lambda \equiv \Delta t_{inf}/\tau_{f},
\label{eq:lambda}
\end{equation}
where $\tau_{f}$ is the characteristic time of flight between two collisions,
which is determined by the density ($\rho$) and the mean velocity of agents 
($\langle v \rangle$). The critical exponents of the transition 
to spreading were presented by us for the same kind of system~\cite{us}, where the study was done in terms of the fraction of infected individuals ($F_{IM}$).
Here we go further and characterize the behavior of the
clusters of infected individuals. When agent $j$ infects agent $i$ 
a link is created among them, the link lasts until one of them heals, 
meanwhile each of them  continues making links with other susceptible
agents through the same rule. 
A cluster is thus defined as a group of infected agents connected by links. 
Note that in contrast to percolation, where clusters are given by occupied 
lattice sites connected by nearest-neighbor distances, for this model each cluster gives a group of agents infected in a given period of time linked by a 
relation of contagion. Isolated infected agents are regarded as clusters
of size unity and any cluster consisting of $s$ connected agents is an
$s-cluster$. We borrow the notation from Stauffer's book 
on percolation theory~\cite{Stauffer} and define here $n_{s}=N_{s}/N$ 
as the number of $s$-clusters per agent, where $N_{s}$ is the number of 
clusters of size $s$ and $N$ the total number of agents in the system. 
For different values of $\lambda$, in the
next section we present the results of the calculation of 
the first three moments of the cluster size distribution. Namely: $\sum_{s} n_{s}$,
$\sum_{s} s n_{s}$, $\sum_{s} s^{2} n_{s}$. Those quantities give us, respectively, 
information about: the total number of clusters, the fraction of
infected agents and the mean size of clusters. In order to keep the analogy
with the calculation on percolation, we sum
over all values of $s$ excluding the largest cluster ($S_{major}$). 
We also present, the  calculations of 
$F_{major}=S_{major}/N$, the fraction of agents that belong to the
largest cluster and $F_{IM}=N_{inf}/N$, the fraction 
of agents that are infected.
\begin{figure}
\unitlength 1mm
\begin{center}
\leavevmode 
{\includegraphics[width=5.75cm]{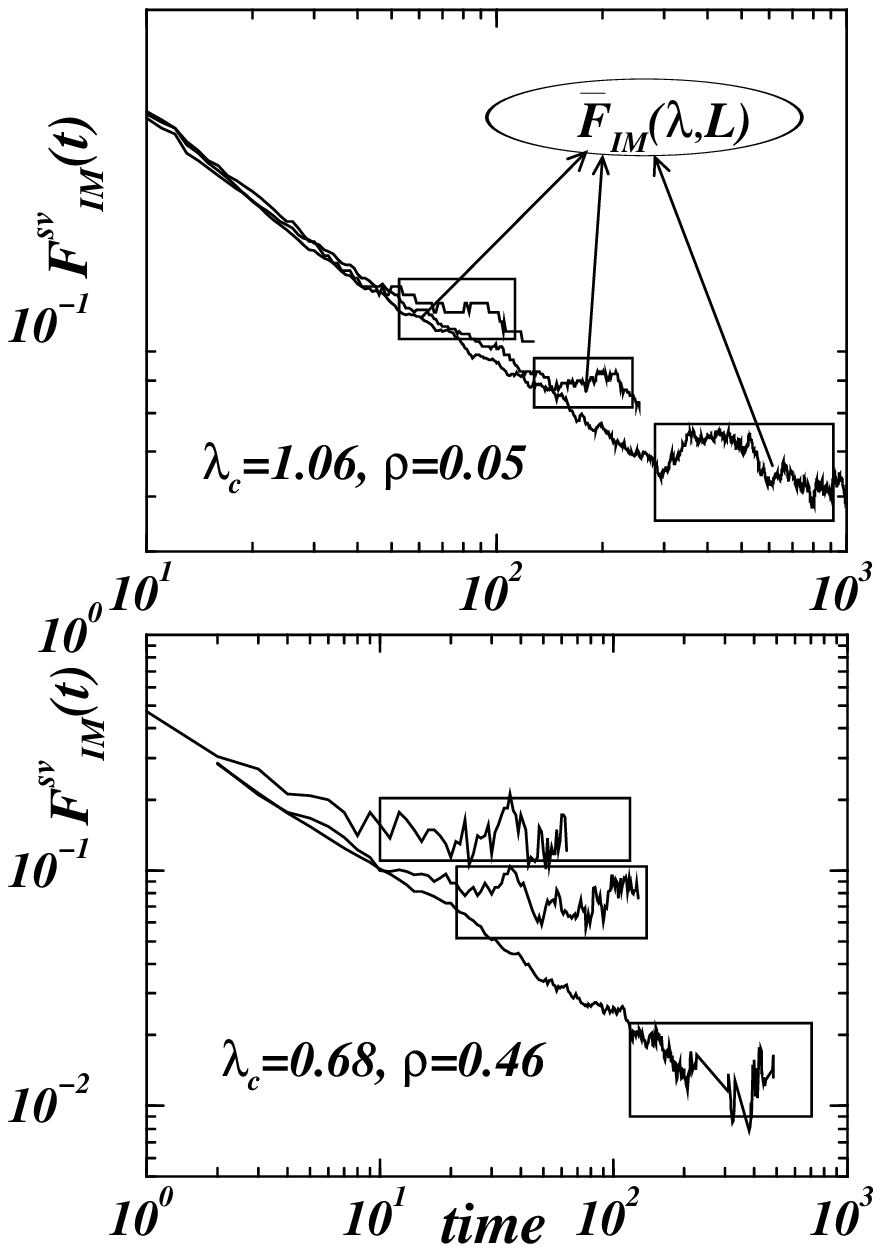}
\includegraphics[width=5.78cm]{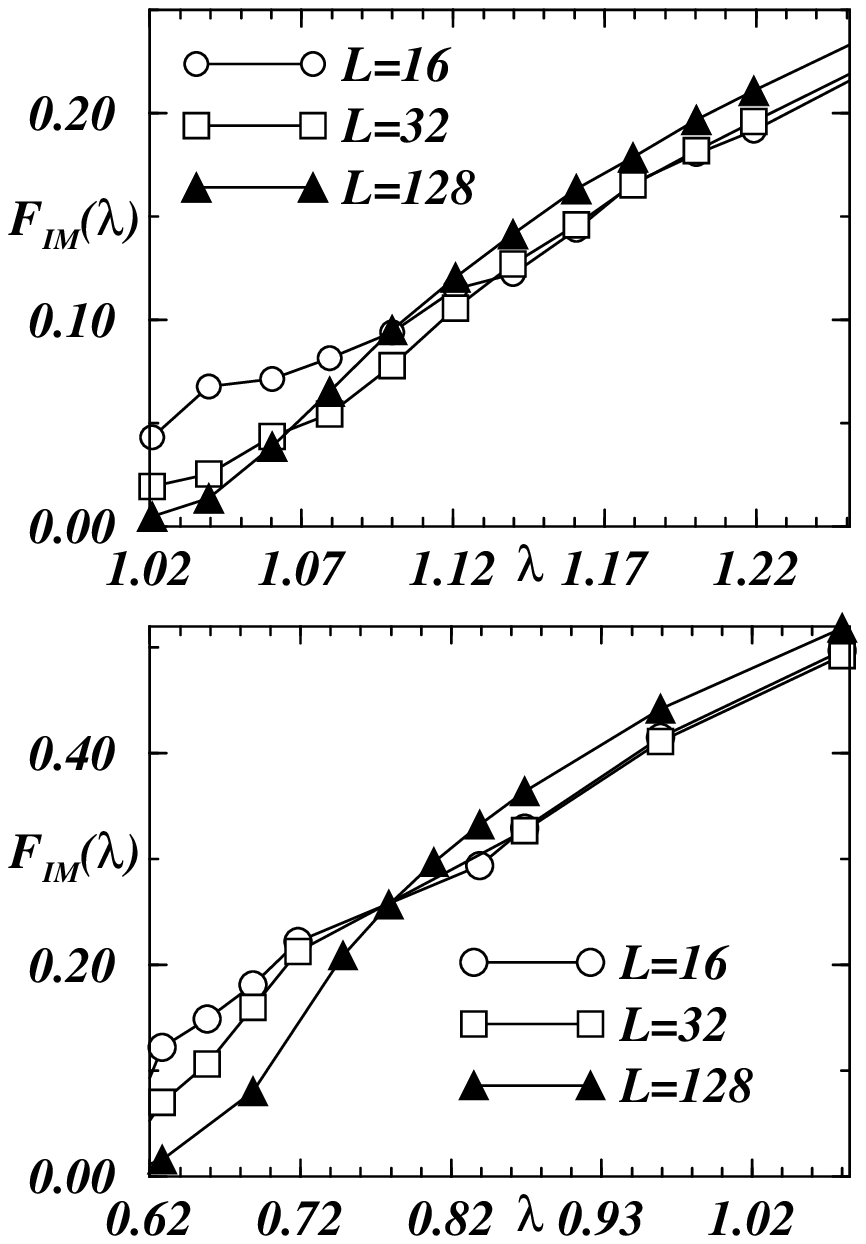}}
\end{center}
\caption{\protect Left: Fraction of infected individuals  from surviving
trials versus time at $\lambda=\lambda_{c}$, starting with half of the
population infected. At the top, the results for $\rho=0.05$ and $\lambda=1.06$
and at the bottom $\rho=0.46$ and $\lambda=0.68$. System sizes $N = 32\times32$, $64\times64$, $128\times128$ 
(from top to bottom). Right: Quasi-stationary fraction of infected agents versus
$\lambda$ for the same densities (Top: $\rho=0.05$. Bottom: $\rho=0.46$).}
\label{fig1}       
\end{figure}
    
\section{Results} 
\begin{figure}
\unitlength 1mm
\begin{center}
\leavevmode
{\includegraphics[width=6.25cm]{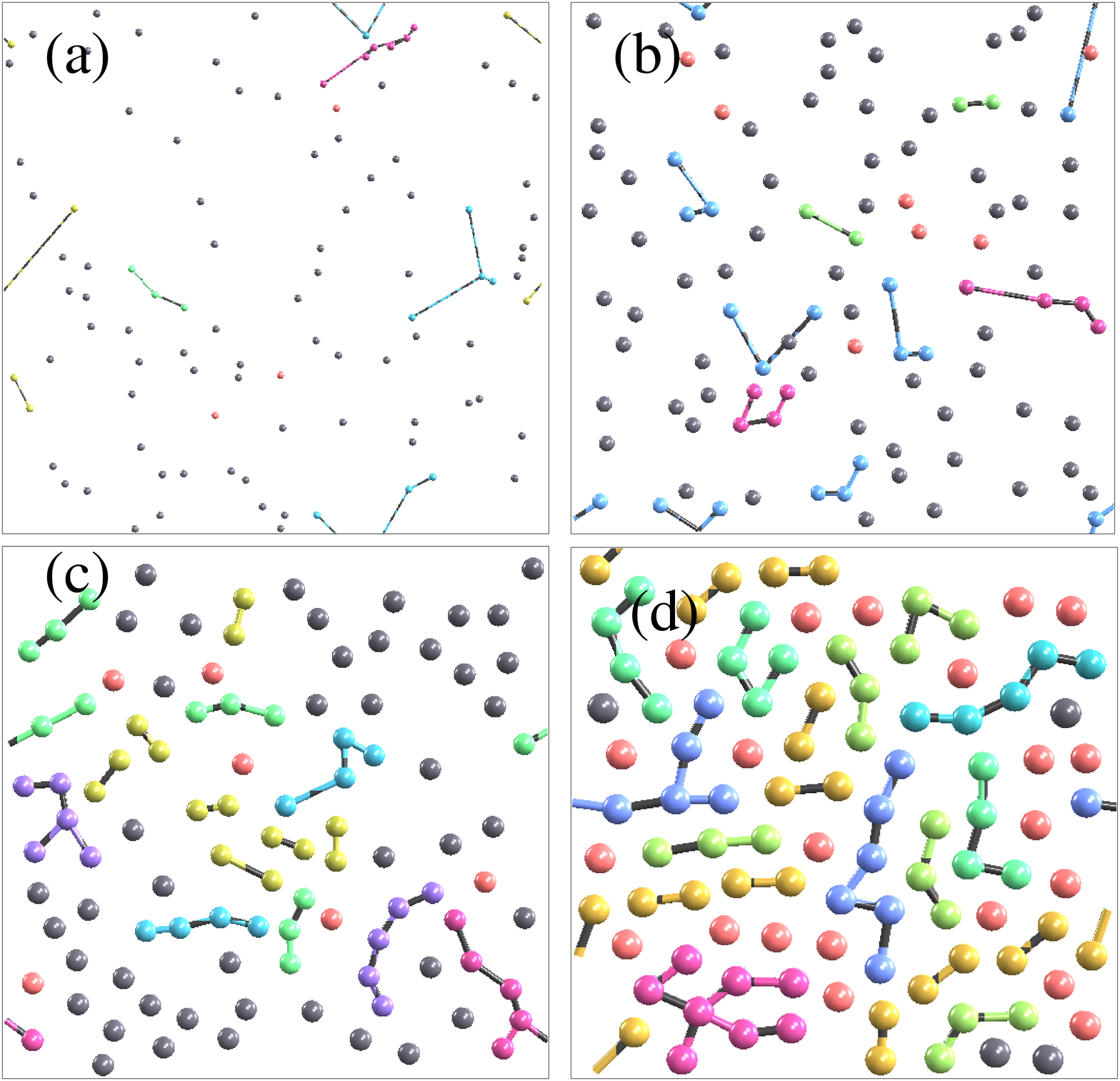}}
{\includegraphics[width=7.5cm]{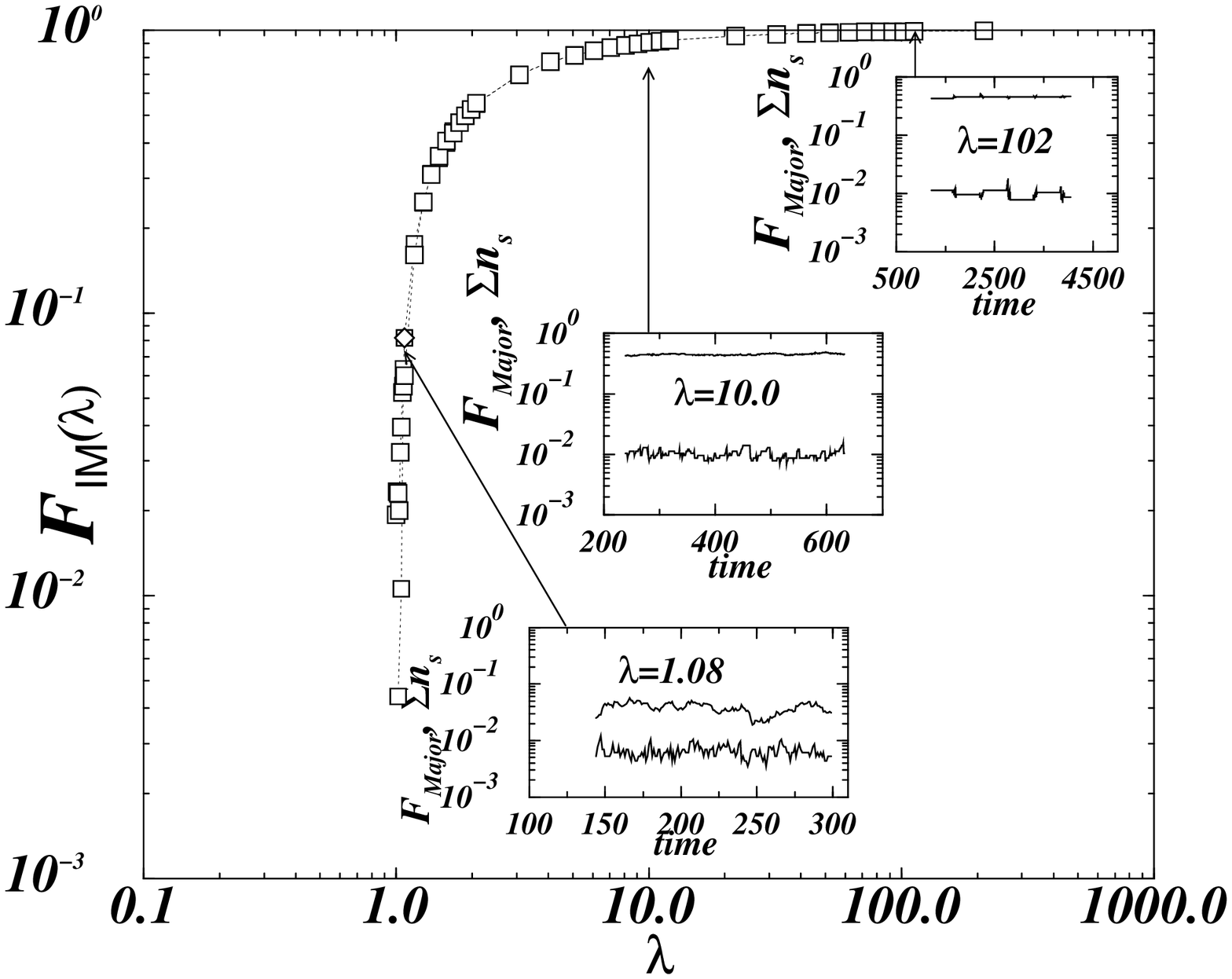}}
\caption{\protect Left: Snapshots of cluster sizes of infected agents for systems
with different densities: (a)$\rho=0.05$, (b)$\rho=0.20$,
(c)$\rho=0.40$ and (d)$\rho=0.80$, in all cases $\lambda=1.5$. 
Right: Quasi-stationary fraction of 
infected agents varying $\lambda$ over three orders of magnitude
(Average over $20$ realizations for $\rho=0.05$ and $N=32\times32$). The insets
show the fraction of infected agents in the largest cluster (lower value)
and the the first moment of the cluster size distribution (upper value) vs. time, at
$\lambda=1.08$, $\lambda=10.0$ and $\lambda=108.0$.
}
\label{fig2}
\end{center}
\end{figure}

For a fixed density, we vary $\lambda$ (Eq.~\ref{eq:lambda})
changing the time of infection ($\Delta t_{inf}$).
Starting with half of the population infected, for 
rate of infections near $\lambda_{c}$, a given trial
may end into the absorbing state after a few time 
steps or it may {\em survive} fluctuating with a 
quasi-stationary fraction of infected agents, marked 
with windows in the left-side of Fig~\ref{fig1}.   
The calculations are made averaging on time at 
the {\em quasi-stationary state}, which is described
by the surviving trials following an initial transient. 
The number of time steps of this transient  
depends on $\lambda$ and on the system size $L$ (see left side
of Fig.~\ref{fig1}). The data here illustrate how 
the mean fraction of infected agents $F^{sv}_{IM}(t)$ (the 
superscript denotes an average restricted to 
surviving trials) approaches its stationary value 
$\bar{F}_{IM}(\lambda,N)$ (in the following, we write
$\bar{F}_{IM}(\lambda,N)$ just like $F_{IM}(\lambda))$. 
In the right side of the same figure we see the graph of 
$F_{IM}(\lambda)$, which becomes 
sharper increasing the system size. We analyze in detail
the number of clusters for the two density values $\rho=0.05$ and
$\rho=0.46$, which have critical rate of spreading
$\lambda_{c}=1.06$ and $\lambda_{c}=0.68$ respectively. Note
that at the critical density $\lambda_c$, surviving trials
tend to stationary values only in the limit $L \rightarrow \infty$.
\\
The left side of Fig.~\ref{fig2} is only for pedagogical reasons, in order
to illustrate how the number of clusters looks in the quasi-stationary state, we see 
snapshots of the clusters of infections for {\em different} systems
densities and the {\em same} rate of infection $\lambda=1.5$, here $N=10\times10$.\\
For $\rho=0.05$ and $\lambda \in [1,200]$, the right side of
Fig.~\ref{fig2} shows the variation of $F_{IM}(\lambda)$ and $N=32\times32$
averaged over $20$ different realizations. 
The insets show the change in time of $F_{major}$ and 
$\sum_{s} n_{s}$, for only one realization with $\lambda=1.08$, 
$\lambda=10.0$ and $\lambda=108$. In contrast to
percolation results, in this model there is no significant variation
of $F_{major}$ with $\lambda$, and the relation
$F_{major} \ll F_{IM}$ remains.
Moreover, the number of clusters  $\sum_{s} n_{s}$ grows
considerably only near $\lambda_{c}$.
\begin{figure}
\unitlength 1mm
\begin{center}
\leavevmode 
{\includegraphics[width=13.0cm]{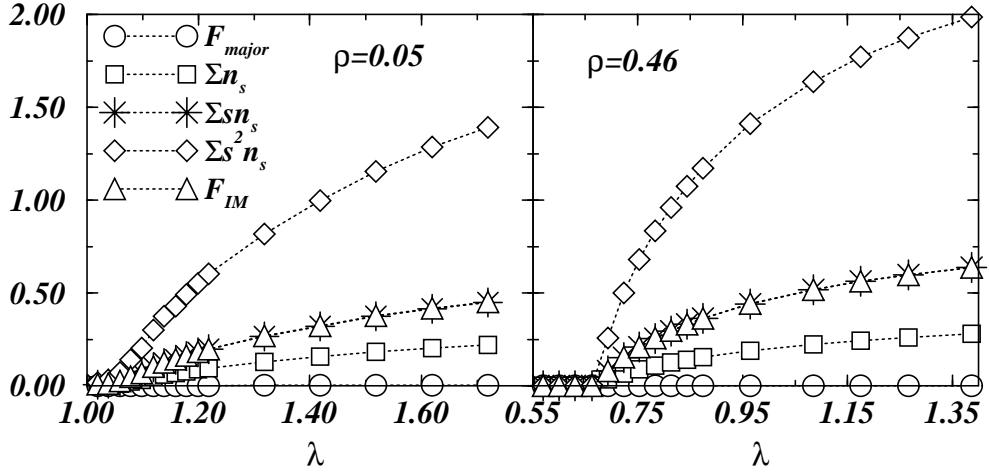}}
\end{center}
\caption{\protect First three moments of the cluster size distribution
, fraction of agents in the largest cluster ($F_{major}$) and
fraction of infected agents ($F_{IM}$) vs. $\lambda$. Average over 50 trials,
system size $N=64\times64$. Left:$\rho=0.05$. Right:$\rho=0.46$}.
\label{fig3}       
\end{figure}

\begin{figure}
\unitlength 1mm
\begin{center}
\leavevmode 
{\includegraphics[width=6.75cm]{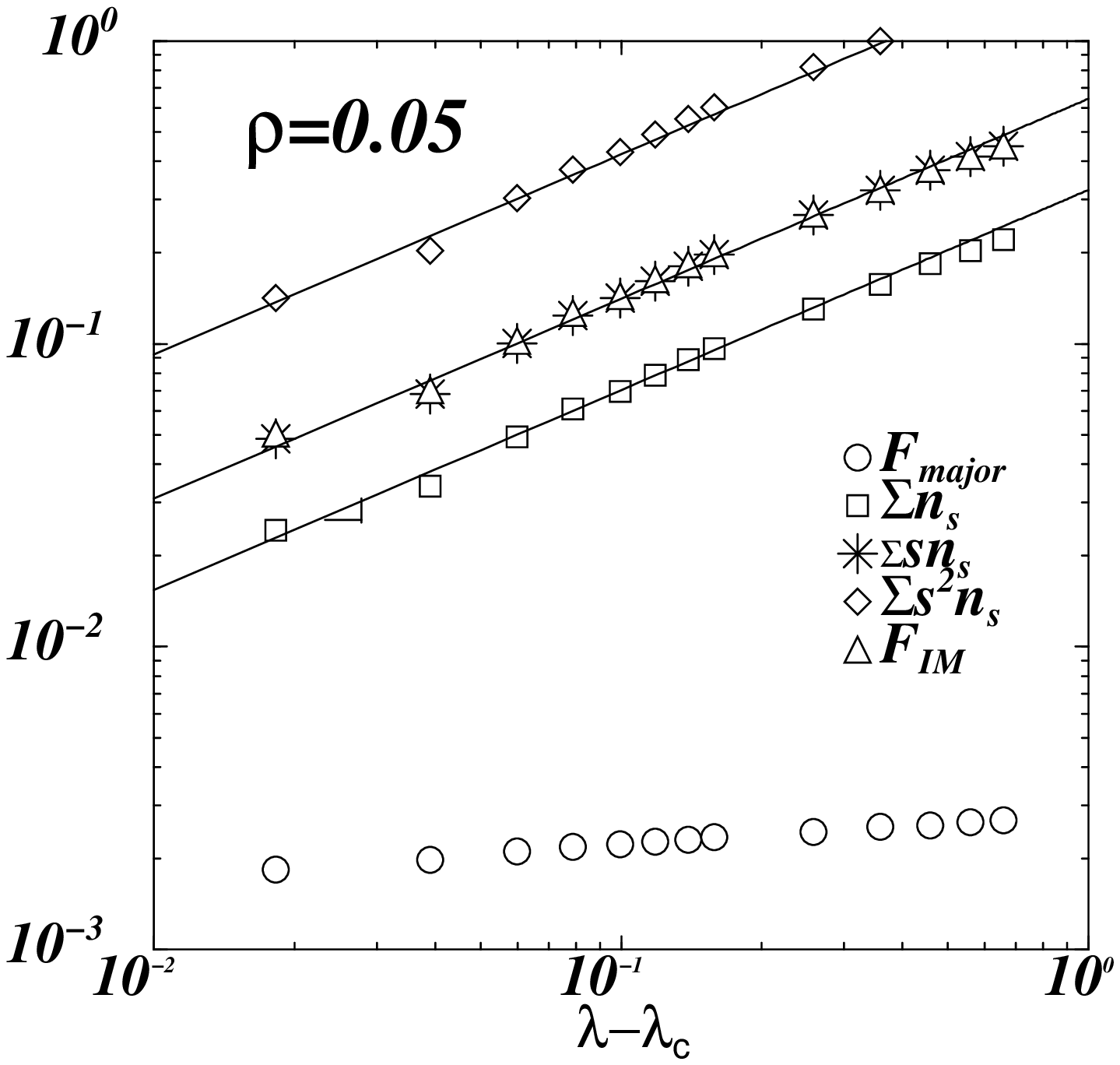}
\includegraphics[width=6.75cm]{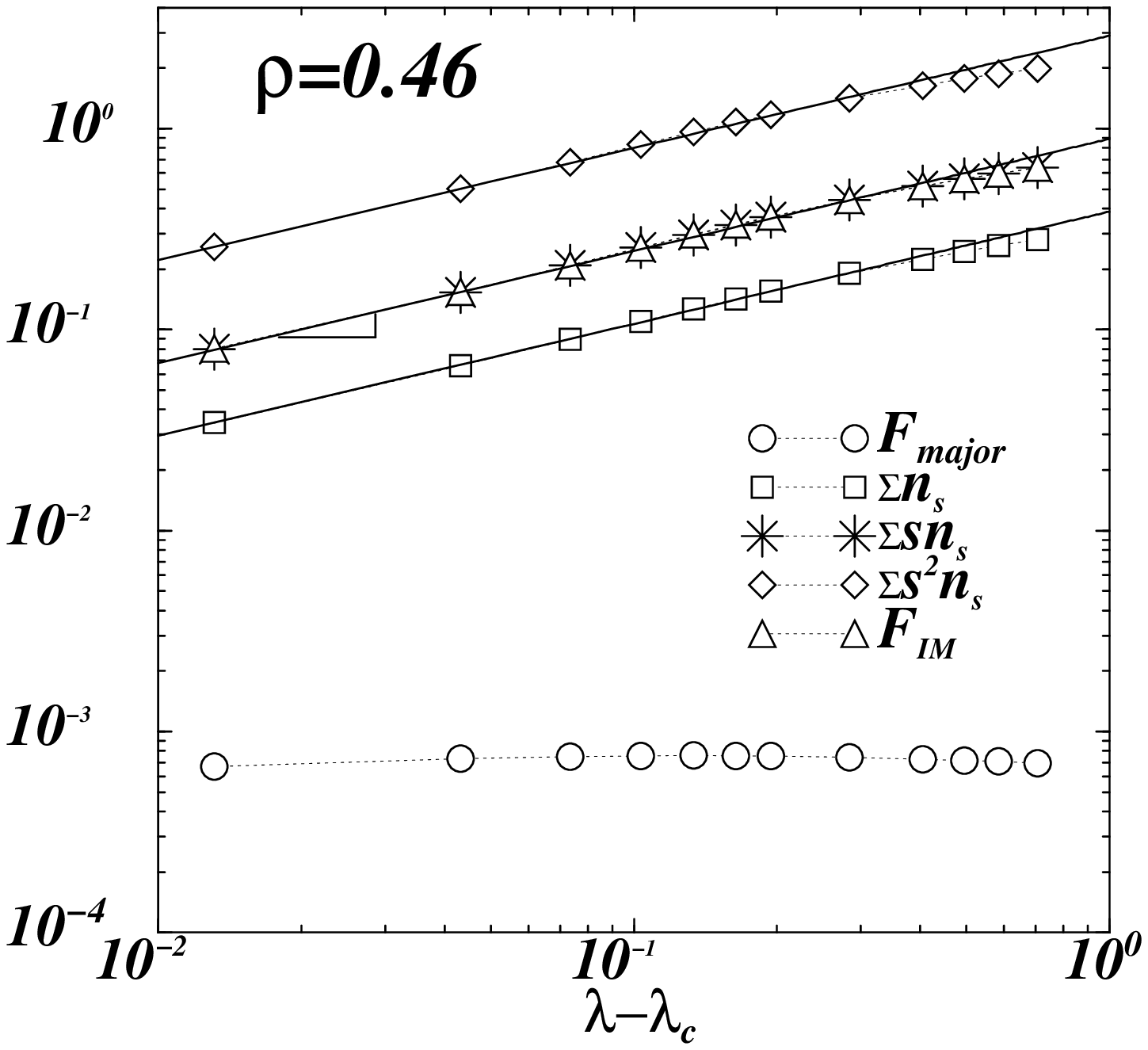}}
\end{center}
\caption{\protect Same results of Fig.~\ref{fig3} plotted
vs. $(\lambda-\lambda_{c})$. The solid lines are regressions
of the form $m_{i}(\lambda-\lambda_{c})^{\beta}$ with
$m_{i}$ the coefficient of the $i${\em th} moment.
Left: $\lambda_{c}=1.06$, $\beta=0.66$,
$m_{0}=0.321$, $m_{1}=2m_{0}$, and $m_{2}=6m_{0}$. 
Right: $\lambda_{c}=0.68$, $\beta=0.56$, $m_{0}=0.386$,
$m_{1}=2.3m_{0}$, and $m_{2}=7.5m_{0}$}
\label{fig4}       
\end{figure}
In Fig.~\ref{fig3} for $\rho=0.05$
and $\rho=0.46$, we plot the behavior of the cluster
numbers near their respective $\lambda_{c}$. As the largest
cluster remains small compared to the total number
of agents ($S_{major} \ll N$), we have $F_{IM}(\lambda) \sim \sum_{s} s n_{s}$.Additionally one can see that
$\sum_{s} s n_{s}$ and $\sum_{s} s^{2}n_{s}$ show the same critical
behavior as $F_{IM}(\lambda)$, plotted in detail in Fig.~\ref{fig4}.
We observe that all the moments of the cluster
size distribution present exactly the same critical behavior
than $F_{IM}$, namely $~(\lambda -\lambda_{c})^{\beta}$, where
$\beta$ depends on the density of the system.  
\section{Conclusions}
This work showed that the cluster size distribution 
of infected individuals is described in terms of the spreading rate 
($\lambda$) and the same exponents ($\beta$) 
previously known for the total mass of infection. Although the agents are free
to move there is a homogeneous size distribution of infected
clusters at the critical rate of infection, and we did not 
find any critical exponent associated with the cluster sizes. 
Comparing with the traditional $SIS$ model on a static
network we confirm that mobility and spatial correlations change
the value of the critical exponent $\beta$ of the fraction of 
infected population, 
and to the same extent the cluster size distribution of infection.

\end{document}